\documentclass{optica-article}

\journal{opticajournal} 

\articletype{Research Article}

\usepackage{xspace}
\newcommand{\MATLAB}{\textsc{Matlab}\xspace}

\newcommand{\dbar}[1]{\overline{\overline{#1}}}

\begin{document}

\title{Perfect all-angle asymmetric transmission via normal susceptibilities: exact spatial derivative by local meta-atoms and nonlocal metasurfaces}

\author{Amit Shaham,\authormark{1,2,$\ddag$} and Ariel Epstein,\authormark{1,*}}

\address{\authormark{1}Andrew and Erna Viterbi Faculty of Electrical and Computer Engineering, Technion---Israel Institute of Technology, Haifa 3200003, Israel\\
\authormark{2}Currently with the Photonics Initiative of the Advanced Science Research Center, City University of New York, New York, New York 10031, USA}

\email{\authormark{$\ddag$}samitsh@campus.technion.ac.il\\ \authormark{*}epsteina@ee.technion.ac.il} 


\begin{abstract*} 
We present a systematic methodology for realizing accurate asymmetric all-angle transmission in nonlocal metasurfaces. As a representative example, we derive closed-form susceptibility conditions for exact first-order spatial differentiation of unity numerical aperture, clarifying the role of each underlying balance. We provide rigorous and detailed designs of physically meaningful structures that directly feature such susceptibilities: a conceptual local meta-atom and a realistic nonlocal multilayered printed circuit board (PCB). Importantly, the latter leverages an intricate system of nearfield coupling beyond standard homogenization. Validated in simulations, our results provide a general and modular route to high-resolution asymmetric nonlocal metasurfaces for optical analog processing.

\end{abstract*}

\section{Introduction}
\label{Sec:Intro}
Metasurfaces (MS) have unfolded an exhilarating era of electromagnetic and optical possibilities, wherein extreme wave scattering can be molded by subwavelength patterning of artificial particles \cite{Glybovski2016,Chen2016,Li2018,Quevedo-Teruel2019}. Beyond its attractive applications of field transformation \cite{Pfeiffer2013Huygens,Monticone2013,Pfeiffer2013Cascaded,Selvanayagam2013,Pfeiffer2013Millimeter,PfeifferNano2014,Wong2014,Asadchy2015Functional,Epstein2016,Asadchy2016,Epstein2016OBMS,Chen2018,Ataloglou2021} and antenna enhancement \cite{Fong2010,Patel2011,Minatti2011,Morote2014,Monticone2015,EpsteinNature2016,Minatti2016Flat,Minatti2016Synthesis,Smith2017,Sanchez2018,Xu2023}, this rich area has driven fundamental advances in wave physics. One such notable advance is the utilization of nonlocal responses in composite materials. 

Unlike local media, which are assumed to respond to an electromagnetic excitation at the same position where applied, nonlocal media react also to the fields at other (distant) locations \cite{Overvig2022,Shastri2023}. In the spectral domain, this additional degree of freedom translates to spatial dispersion, i.e., wavevector or angular sensitivity, which often stands as a central characteristic to be regulated---whether leveraged or suppressed---during design. As so, recent trends have sought to control and perfect this feature for a new generation of devices, including optical analog computers \cite{Silva2014,Zhu2017,Guo2018,Kwon2018,Momeni2019,Cordaro2019,Abdolali2019,Kwon2020,Abdollahramezani2020,Zangeneh2021,Momeni2021,He2022Review,Cordaro2023,Nikkhah2024,Hu2024} for rapid processing of fields and images without digitization and power consumption; space-squeezing plates (spaceplates) \cite{Guo2020,Chen2021,Reshef2021,Page2022,Shastri2022,DelMastro2022,Diaz-Fernandez2024} for emulating long-range free-space propagation within compact structures and thus miniaturizing optical systems; transformations beyond local power conservation \cite{Epstein2016PRL,Kwon2017,Tcvetkova2018,DHKwon2018,Ataloglou2021Arbitrary,Xu2022}; and advanced leaky-wave devices \cite{Xu2023}.

\begin{figure}
    \includegraphics[width=\textwidth]{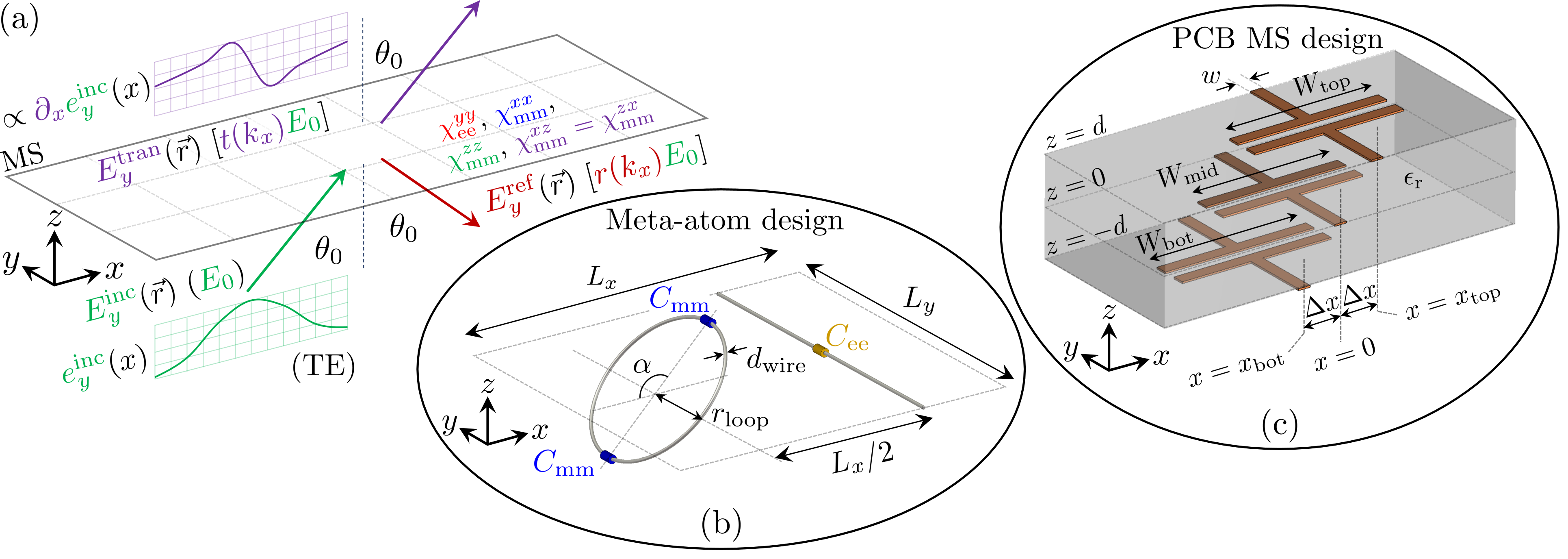}
    \caption{(a) Physical configuration of a spatially deriving (asymmetric) MS: the spatial profile of the incident field $e_{y}^{\mathrm{inc}}(x)$ is derived through transmission $\propto\partial_{x}e_{y}^{\mathrm{inc}}(x)$. In the spectral domain, the incident plane wave [$E_{y}^{\mathrm{inc}}(\vec{r})$] of tangential wavenumber $k_{x}$ is specularly reflected [$E_{y}^{\mathrm{ref}}(\vec{r})$] and directly transmitted [$E_{y}^{\mathrm{tran}}(\vec{r})$] with the desired transmission coefficient $t(k_{x})\propto -jk_{x}$. (b) All-angle meta-atom design (one period) consisting of a subwavelength loaded loop rotated by angle $\alpha$ around the $y$-axis and a loaded line (Sec.\ \ref{Subsec:MetaAtom}). (c) Realistic all-angle trilayered PCB design (one period) of misaligned printed loaded strips (Sec. \ref{Subsec:PCB}). Finalized parameter values are provided in the main text as a part of their careful tuning process.
}
\label{Fig:Setup}
\end{figure}

Despite the fruitful demonstrations above, other major aspects of nonlocal MSs---particularly, systematic and accurate shaping of their response over the entire angular (propagating wavevector) domain---still lack rigorous and modular engineering schemes; this often results in undesired performance suboptimality and limitations. As an instructive example for this issue, consider the specific nonlocal functionality of first-order spatial differentiation along the $x$-direction [$\partial_{x}$, Fig.\ \ref{Fig:Setup}(a)] \cite{Silva2014,Zhu2017,Kwon2018,Momeni2019,Cordaro2019,Abdolali2019,Kwon2020,Momeni2021}, defined by the wavevector-dependent transmission coefficient $t(k_{x})\propto -jk_{x}$ ($k_x$ is the $x$-directed wavenumber of a plane wave with $x$-dependence of $e^{-jk_x x}$). Beyond its mathematical importance to optical analog processing, this gradient operation can be used as a practical tool for edge detection and sharpening of images, as its high-pass characteristics filter out slowly varying paraxial components while emphasizing the rapidly varying spectral contents. From yet another fundamental perspective, first-order derivative can be viewed as an archetype of asymmetric transmission with respect to the optical axis, $t(-k_{x})\neq t(k_{x})$, which generally necessitates meta-atoms of broken symmetry along both the transverse and longitudinal axes of the MS \cite{Kwon2018,Momeni2019,Abdolali2019,Cordaro2019,Achouri2020,Shastri2023,Abouelatta2025}; this introduces additional intricacy to be handled during design.

To realize such a spatially differentiating device in practice, one must propose a realistic meta-atom inclusion that achieves the $t(k_{x})\propto -jk_{x}$ goal when arranged periodically on the MS plane. The maximal range of transverse wavenumbers $k_{x}$ over which the realized transmission follows the ideal one defines the numerical aperture (NA) of the apparatus. To maximize resolution and mitigate undesired distortions, the MS inclusions should be carefully chosen and tuned such that as high an NA as possible is to be achieved.

However, lacking a systematic methodology for shaping the angular dependence of transmission \emph{directly}, past approaches have resorted to either ``off-the-shelf'' or ``ad-hoc'' inclusions, such as loaded loops and wires \cite{Kwon2018} or structured dielectric or metallic constructs \cite{Momeni2019,Cordaro2019,Abdolali2019,Kwon2020,Momeni2021}, modified according to certain first principles---mainly the aforementioned symmetry-breaking requisite. Once the generic geometry had thus been determined, its features were next optimized based on certain performance metrics, mostly indirect, for instance, the loci and coupling strengths of the underlying Fano resonances. Consequently, although the realized responses behaved desirably near the optical axis (i.e., around the normal incidence of $k_{x}=0$), their accuracy rapidly deteriorated for larger transverse wavenumbers, resulting in modest NAs (typically fewer than ten degrees). Indeed, for the purposes of processing paraxial images of visible light with feature sizes perceivable by the naked human eye, the NA performance of these devices may be sufficient. Nevertheless, other applications, for instance, microscopy, require substantially larger resolutions, which demand significant NA improvement in such nonlocal MSs.

Very recently, we have addressed this acceptance-angle challenge by proposing an alternative framework for shaping the angular response of MSs \emph{directly} and for \emph{all} the (propagating) angles of incidence \cite{Shaham2024AdvOptMat,Shaham2025IEEETAP}. By rigorous collocation of tangentially and normally polarizable \emph{local} electromagnetic responses, we have shown that the spatial dispersion of the MS scattering can be tightly controlled, particularly near the subtle grazing angle---the key gateway to unity NA. Furthermore, we have comprehensively revealed how such delicate mixtures of tangential and normal susceptibilities can be implemented through practical multilayered printed-circuit-boards (PCBs) by carefully tuning the \emph{nonlocal} mechanisms of multiple reflections at play. Although this new methodology has established a realistic path to all-angle nonlocal MSs on demand, its use is currently limited to scenarios where the transmission properties are symmetric around the optical axis, $t(-k_{x})=t(k_{x})$. The reason is that the standard surface-impedance (transmission-line) models of such PCBs \cite{Pfeiffer2013Cascaded,Pfeiffer2013Millimeter,Epstein2016,Epstein2016OBMS,Epstein2016PRL,Chen2018,Ataloglou2021,Shaham2024AdvOptMat,Shaham2025IEEETAP} are subject to homogenization in a way that each layer interacts with the other layers only through the propagating fundamental diffracted order; the $k_{x}$-symmetry is then inherited from that of the impedance-sheet responses. Indeed, several fundamental operation modes, e.g., the all-angle-transparent radome [generalized Huygens' condition, $t(k_{x})\equiv 1$] and all-angle perfect-magnetic-conductor [PMC, $r(k_{x})\equiv 1$ reflection] in \cite{Shaham2024AdvOptMat,Shaham2025IEEETAP}, are supported as they satisfy this symmetry restriction; however, other asymmetric applications with $t(-k_{x})\neq t(k_{x})$, for instance, spatial differentiators, are presently excluded.

In this paper, we demonstrate how our TE all-angle nonlocal MS framework in \cite{Shaham2025IEEETAP} can be extended to support such asymmetric transmission scenarios in closed form by introducing an additional tangential-normal anisotropic susceptibility component and modifying the PCB layout accordingly. Following the standard generalized sheet transition conditions (GSTCs) \cite{Idemen1990,Tretyakov2003,Kuester2003,Achouri2015}, we first express the general asymmetric TE MS scattering versus angle that supports a versatile range of all-angular behaviors (Sec.\ \ref{Subsec:GeneralASymmetric}). Thereafter, we focus on the particular asymmetric functionality of perfect all-angle spatial differentiation as a representative (and practically appealing) example to demonstrate the practical design procedure. In particular, we derive the closed-form condition for perfect all-angle spatial differentiation in terms of local surface susceptibilities and highlight its properties (Sec.\ \ref{Subsec:SpatialDiff}) \cite{Shaham2024APSURSI}.
Next, inspired by the rigorous meta-atom validation of the generalized Huygens' condition in \cite{Shaham2024AdvOptMat}, we propose an electromagnetically justified meta-atom configuration of loaded loop and wire [Fig.\ \ref{Fig:Setup}(b)] to validate the all-angle spatially deriving MS condition in full-wave simulations (Sec.\ \ref{Subsec:MetaAtom}) \cite{Shaham2024APSURSI}. Finally, building on our previous findings in \cite{Shaham2025IEEETAP} that thin PCB cascades of printed loaded strips can universally realize normal susceptibilities for all angles, we demonstrate how carefully breaking their symmetry through misalignment can further emulate the additional anisotropic component through asymmetric \emph{nonlocal} nearfield interactions. Specifically, violating the aforementioned homogenized regime through higher-order nearfield coupling and following rigorous analytical models of multilayered strip arrays \cite{Wait1955,Tretyakov2003,Rabinovich2017,Molero2017,Rabinovich2018Analytical,Rabinovich2018Arbitrary,Xu2021,Yashno2023}, we implement a PCB (TE) spatial differentiator [Fig.\ \ref{Fig:Setup}(c)] with a highly accurate simulated response of near-unity NA, drastically surpassing the above past implementations in this regard (Sec.\ \ref{Subsec:PCB}). Overall, this work provides a systematic and universal scheme to perfect the accuracy and NA of realistic MSs for high-resolution nonlocal field manipulation, also for demanding asymmetric scenarios in optical analog processing.

\section{Theory}

\subsection{General asymmetric scattering off TE MSs with local susceptibilities}
\label{Subsec:GeneralASymmetric}
We consider a free-space-surrounded homogeneous MS situated on the $z=0$ plane [Fig.\ \ref{Fig:Setup}(a)]. This MS supports induced electric ($\vec{P}_{\mathrm{s}}$) and magnetic ($\vec{M}_{\mathrm{s}}$) surface polarizations that, in turn, generate field discontinuities governed by the GSTCs \cite{Idemen1990,Tretyakov2003,Kuester2003,Achouri2015}
\begin{equation}
\label{Eq:GSTCs}
    \begin{aligned}
    \hat{z}\times\left( \vec{H}_{\mathrm{t}}^{+}-\vec{H}_{\mathrm{t}}^{-} \right)&=j\omega\vec{P}_{\mathrm{st}}-\hat{z}\times\vec{\nabla}_{\mathrm{t}}M_{\mathrm{s}z},\\
    \left( \vec{E}_{\mathrm{t}}^{+}-\vec{E}_{\mathrm{t}}^{-}\right)\times\hat{z}&=j\omega\mu_{0}\vec{M}_{\mathrm{st}}-\vec{\nabla}_{\mathrm{t}}\left( \frac{P_{\mathrm{s}z}}{\epsilon_{0}}\right)\times\hat{z},
    \end{aligned}
\end{equation}
where $\hat{z}$ is the normal to the MS plane; ``$\mathrm{t}$'' subscript denotes vector components tangential to the MS plane; $\pm$ superscripts respectively denote field values near the top ($z\to 0^{+}$) and bottom ($z\to 0^{-}$) facets; $\epsilon_0$ and $\mu_0$ are the permittivity and permeability of free space; and global harmonic time dependence of $e^{j\omega t}$ is assumed and suppressed.

Induced by the incident fields, these surface polarization distributions along the MS ($\vec{P}_{\mathrm{s}}$ and $\vec{M}_{\mathrm{s}}$) are fully determined by the electromagnetic properties and geometric features of its subwavelength constituents. Specifically, for local (and linear) MS inclusions that respond only to the fields applied in situ, the polarizations are driven by the averaged fields, $\vec{E}^{\mathrm{av}}=\frac{1}{2}\left(\vec{E}^{+}+\vec{E}^{-}\right)$ and $\vec{H}^{\mathrm{av}}=\frac{1}{2}\left(\vec{H}^{+}+\vec{H}^{-}\right)$, via \cite{Idemen1990,Tretyakov2003,Kuester2003,Achouri2015}
\begin{equation}
\label{Eq:Suscept}
    \begin{aligned}
		\vec{P_{\mathrm{s}}}&=\epsilon_{0}\dbar{\chi}_{\mathrm{ee}}\cdot\vec{E}^{\mathrm{av}}+c^{-1}\dbar{\chi}_{\mathrm{em}}\cdot\vec{H}^{\mathrm{av}},\\	\vec{M}_{\mathrm{s}}&=\eta_{0}^{-1}\dbar{\chi}_{\mathrm{me}}\cdot\vec{E}^{\mathrm{av}}+\dbar{\chi}_{\mathrm{mm}}\cdot\vec{H}^{\mathrm{av}},
	\end{aligned}
\end{equation}
where $\dbar{\chi}_{\mathrm{ee}}$, $\dbar{\chi}_{\mathrm{mm}}$, $\dbar{\chi}_{\mathrm{em}}$, and $\dbar{\chi}_{\mathrm{me}}$ are the surface electric, magnetic, (bianisotropic) electromagnetic, and (bianisotropic) magnetoelectric susceptibility tensors, respectively.

We next illuminate the MS from below ($z<0$) by a $y$-polarized TE plane wave expressed via $E_{y}^{\mathrm{inc}}(\vec{r})=E_{0}e^{-j(k_{x}x+k_{z}z)}$, whose tangential and normal wavenumbers, $k_{x}=k_0\sin\theta_0$, and $k_{z}=k_0\cos\theta_0$, are related to the angle of incidence $\theta_0$ through the wavenumber in free space $k_0=\omega/c$, where $c=(\mu_0 \epsilon_0)^{-1/2}$ is the speed of light in free space. Since the macroscopic MS constituents are independent of the transverse coordinates, the scattered waves include only the specular reflection and direct transmission with angles (tangential wavenumbers) of departure identical to that of the incident wave $\theta_0$ ($k_{x}=k_0\sin\theta_0$), as illustrated Fig.\ \ref{Fig:Setup}(a).

Herein, we focus on $y$-polarized TE MSs ($E_{x}=E_{z}=H_{y}\equiv 0$, $\partial_{y}\equiv 0$), such that the above scattered waves of specular reflection, $E_{y}^{\mathrm{ref}}(\vec{r})$, and direct transmission, $E_{y}^{\mathrm{tran}}(\vec{r})$, remain $y$-polarized; overall, the scattered fields can be expressed via
\begin{equation}
\label{Eq:EyScat}
    \begin{aligned}
        E_{y}^{\mathrm{ref}}(\vec{r})&=r(k_{x})E_0e^{-j\left(k_{x}x-k_{z}z\right)}&(z<0),\\
        E_{y}^{\mathrm{tran}}(\vec{r})&=t(k_{x})E_0e^{-j\left(k_{x}x+k_{z}z\right)}&(z>0),
    \end{aligned}
\end{equation}
where $r(k_{x})$ and $t(k_{x})$ define the wavevector-dependent reflection and transmission coefficients off the MS in Fig.\ \ref{Fig:Setup}(a). The most general subset of local susceptibility configurations that supports such TE-only MS operation without polarization conversion reads \cite{Achouri2018,Momeni2019,Abdolali2019,Momeni2021}
\begin{equation}
\label{Eq:SusceptComponents}
    \begin{aligned}
    &\dbar{\chi}_{\mathrm{ee}}=
    \begin{bmatrix}
        0 &0  &0\\
        0   &\chi_{\mathrm{ee}}^{yy}    &0\\
        0   &0   &0
    \end{bmatrix}
    ,
    &\dbar{\chi}_{\mathrm{mm}}=
    \begin{bmatrix}
        \chi_{\mathrm{mm}}^{xx} &0  &\chi_{\mathrm{mm}}^{xz}\\
        0   &0    &0\\
        \chi_{\mathrm{mm}}^{zx}   &0   &\chi_{\mathrm{mm}}^{zz}
    \end{bmatrix},\\
    &\dbar{\chi}_{\mathrm{em}}=
    \begin{bmatrix}
        0 &0  &0\\
        \chi_{\mathrm{em}}^{yx}   &0    &\chi_{\mathrm{em}}^{yz}\\
        0   &0   &0
    \end{bmatrix}, &\dbar{\chi}_{\mathrm{me}}=
    \begin{bmatrix}
        0 &\chi_{\mathrm{me}}^{xy}  &0\\
        0   &0    &0\\
        0   &\chi_{\mathrm{me}}^{zy}   &0
    \end{bmatrix}.
    \end{aligned}
\end{equation}
Furthermore, to enable passive designs without external bias, we enforce reciprocity \cite{Achouri2015,Achouri2018,Momeni2019,Abdolali2019}, which imposes $\chi_{\mathrm{mm}}^{zx}=\chi_{\mathrm{mm}}^{xz}$, $\chi_{\mathrm{me}}^{xy}=-\chi_{\mathrm{em}}^{yx}$, and $\chi_{\mathrm{me}}^{zy}=-\chi_{\mathrm{em}}^{zy}$; keeping these relations in mind, we shall often express the following analysis only in terms of $\chi_{\mathrm{mm}}^{xz}$, $\chi_{\mathrm{em}}^{yx}$, and $\chi_{\mathrm{em}}^{yz}$ for brevity.

As shown also in \cite{Momeni2019,Abdolali2019,Momeni2021}, substituting Eqs.\ (\ref{Eq:Suscept})--(\ref{Eq:SusceptComponents}) in the GSTCs (\ref{Eq:GSTCs}) yields the dependence of the scattering coefficients on the wavevector (angle),
\begin{equation}
\label{Eq:ScatCoeffs}
    \begin{aligned}
        r(k_x)&=\frac{r_{0}+r_{1}\widetilde{k}_{z}+r_{2}\widetilde{k}_{z}^{2}}{d_{0}+d_{1}\widetilde{k}_{z}+d_{2}\widetilde{k}_{z}^{2}+d_{3}\widetilde{k}_{z}^{3}},&
        t(k_{x})&=\frac{j\widetilde{k}_{z}\left(t_0+t_{1}\widetilde{k}_{x}+t_{2}\widetilde{k}_{x}^{2}\right)}{d_0+d_{1}\widetilde{k}_{z}+d_{2}\widetilde{k}_{z}^{2}+d_{3}\widetilde{k}_{z}^{3}},
    \end{aligned}
\end{equation}
where
\begin{equation}
\label{Eq:NormalizedWavenumbers}
    \begin{aligned}
        \widetilde{k}_{x}&=\frac{k_{x}}{k_0}=\sin\theta_0,  &\widetilde{k}_{z}=\frac{k_{z}}{k_0}=\cos\theta_0=\sqrt{1-\widetilde{k}_{x}^{2}}
    \end{aligned}
\end{equation}
are the normalized wavevector components (with respect to the free-space wavenumber $k_0$);
\begin{equation}
\label{Eq:d_coeffs}
    \begin{aligned}
    d_0&=2(\widetilde{\chi}_{\mathrm{ee}}^{yy}+\widetilde{\chi}_{\mathrm{mm}}^{zz}), & d_{1}&=j\left[(\widetilde{\chi}_{\mathrm{ee}}^{yy}+\widetilde{\chi}_{\mathrm{mm}}^{zz})\widetilde{\chi}_{\mathrm{mm}}^{xx}+(\widetilde{\chi}_{\mathrm{em}}^{yx})^{2}-(\widetilde{\chi}_{\mathrm{mm}}^{xz})^{2}-4\right],\\ d_{2}&=2(\widetilde{\chi}_{\mathrm{mm}}^{xx}-\widetilde{\chi}_{\mathrm{mm}}^{zz}), &
    d_{3}&=j\left[(\widetilde{\chi}_{\mathrm{mm}}^{xz})^{2}-\widetilde{\chi}_{\mathrm{mm}}^{xx}\widetilde{\chi}_{\mathrm{mm}}^{zz}\right]
    \end{aligned}
\end{equation}
are the respective coefficients of the $\widetilde{k}_{z}$-polynomial in the denominator common to the reflection and transmission coefficients, $r(k_{x})$ and $t(k_{x})$;
\begin{equation}
\label{Eq:r_coeffs}
    \begin{aligned}
    r_0&=-d_{0}=-2(\widetilde{\chi}_{\mathrm{ee}}^{yy}+\widetilde{\chi}_{\mathrm{mm}}^{zz}), & r_{1}&=4\widetilde{\chi}_{\mathrm{em}}^{yx}, & r_{2}=2(\widetilde{\chi}_{\mathrm{mm}}^{xx}+\widetilde{\chi}_{\mathrm{mm}}^{zz})
    \end{aligned}
\end{equation}
are the respective coefficients of the $\widetilde{k}_{z}$-polynomial in the numerator of $r(k_{x})$ [Eq.\ (\ref{Eq:ScatCoeffs})];
\begin{equation}
\label{Eq:t_coeffs}
    \begin{aligned}
    t_0&=-\left[\widetilde{\chi}_{\mathrm{ee}}^{yy}\widetilde{\chi}_{\mathrm{mm}}^{xx}+(\widetilde{\chi}_{\mathrm{em}}^{yx})^{2}+4\right], &
    t_{1}&=-4j\widetilde{\chi}_{\mathrm{mm}}^{xz}, &
    t_{2}&=(\widetilde{\chi}_{\mathrm{mm}}^{xz})^{2}-\widetilde{\chi}_{\mathrm{mm}}^{xx}\widetilde{\chi}_{\mathrm{mm}}^{zz}=-jd_{3}
    \end{aligned}
\end{equation}
are the respective coefficients of the $\widetilde{k}_{x}$-polynomial in the numerator of $t(k_{x})$; and all the susceptibilities are normalized to dimensionless values via $\widetilde{\chi}=k_0\chi$.

Equations (\ref{Eq:ScatCoeffs})--(\ref{Eq:t_coeffs}) already bear some fundamental insights that are worth mentioning \cite{Momeni2019,Abdolali2019,Momeni2021}. First, the reflection coefficient is symmetric to the transverse inversion $\widetilde{k}_{x}\to-\widetilde{k}_{x}$, i.e., $r(-k_{x})=r(k_{x})$ due to reciprocity (it can be formulated solely in terms of $\widetilde{k}_{z}$, which is an even function of $\widetilde{k}_{x}$); in this regard, the contributions of the reciprocal tangential-normal bianisotropic components $\widetilde{\chi}_{\mathrm{me}}^{zy}=-\widetilde{\chi}_{\mathrm{em}}^{yz}$ to the scattering properties cancel one other, rendering this degree of freedom practically redundant. However, the transmission coefficient $t(k_{x})$ features also an odd part [the $t_{1}\widetilde{k}_{x}$ term in its numerator, Eq.\ (\ref{Eq:ScatCoeffs})], which is nonvanishing only if $t_{1}\neq 0$, i.e., when the tangential-normal magnetic component $\widetilde{\chi}_{\mathrm{mm}}^{xz}\neq 0$ is introduced [Eq.\ (\ref{Eq:t_coeffs})]. Note that the asymmetric-transmission resulting from this term does not violate reciprocity (see \cite{Pfeiffer2016,Kwon2018,Achouri2020,Shastri2023}). Therefore, for reciprocal MSs, polarization-preserving asymmetric scattering with respect to the transverse wavenumber, e.g., odd-order spatial differentiation, can be accomplished only in transmission \cite{Momeni2019,Abdolali2019,Momeni2021}.

With these scattering properties at hand, it is possible to control the angular response of the MS by adequately tuning the susceptibility values. To achieve a specific desired operation---especially an asymmetric one, for instance, spatial differentiation [$t(k_{x})\propto-jk_{x}$]---a recent line of reports \cite{Momeni2019,Abdolali2019,Momeni2021} has suggested optimizing the values of these susceptibilities according to certain angularly weighted least-squared-error criteria. As discussed in Sec.\ \ref{Sec:Intro}, while such approaches yield commendable accuracies along several tens of degrees around the optical axis, they still present several gaps that require addressing. First, such optimization schemes are not guaranteed to reach the global maximum of NA achievable by this class of MSs---neither for spatial differentiation nor for more general functionalities. This may deprive full exploitation of the potentially attainable NA, resulting in suboptimal resolution performance. Second, thus far, no direct transition has been performed from the abstract goal susceptibility values to the physical meta-atom inclusion: the choice of meta-atom geometry has still been based purely on symmetry-breaking considerations with no rigorous relation to the abstract susceptibility values they are meant to realize. Furthermore, despite providing valuable proof-of-concept in simulations, the meta-atoms proposed in such past work consist mainly of nontrivial geometric shapes suspended in air and are therefore quite challenging to fabricate. In what follows, we show how these issues can be overcome by adopting an alternative design procedure based on our rigorous methodology and insights in \cite{Shaham2024AdvOptMat,Shaham2025IEEETAP}. We shall do so by demonstrating and validating the pivotal asymmetric example of a perfect all-angle spatially deriving MS in detail.

\subsection{Closed-form condition for perfect all-angle spatial differentiation (TE)}
\label{Subsec:SpatialDiff}
To obtain a perfect all-angle spatial differentiation ($|\theta_0|<90^{\circ}$, $|k_{x}|<1$, for the TE polarization), our objective is to find a set of adequate susceptibility values that, when substituted in Eqs.\ (\ref{Eq:ScatCoeffs})--(\ref{Eq:t_coeffs}), would yield a specific angular dependence of scattering proportional to $-j\widetilde{k}_{x}$. As found in \cite{Shaham2024AdvOptMat,Shaham2025IEEETAP}, the most crucial specification to be treated in order to achieve all-angle operation is the scattering behavior at the grazing angles $\theta_0\to\pm90^{\circ}$, which translates, according to Eq.\ (\ref{Eq:NormalizedWavenumbers}), to $\widetilde{k}_{x}\to\pm1$ and $\widetilde{k}_{z}\to0$. In view of Eq.\ (\ref{Eq:ScatCoeffs}), if $d_0=-r_0\neq 0$, then the transmission inevitably drops to $t(k_{x})\to 0$ and the reflection becomes total, $r(k_{x})\to-1$, for grazing incidence scenarios. This behavior is extremely undesired for high-NA transmissive applications, particularly for all-angle spatial differentiation that requires maximal transmission near the grazing angle.

The only way to overturn this limitation is to enforce the recently reported grazing-angle Huygens' condition \cite{Shaham2024AdvOptMat,Shaham2025IEEETAP}, namely, $d_0=-r_0=0$, which yields [Eqs.\ (\ref{Eq:d_coeffs}) and (\ref{Eq:r_coeffs})]
\begin{equation}
\label{Eq:GrazingHuygens}
    \widetilde{\chi}_{\mathrm{mm}}^{zz}=-\widetilde{\chi}_{\mathrm{ee}}^{yy}.
\end{equation}
In analogy to the standard Huygens' condition at normal incidence ($\widetilde{\chi}_{\mathrm{mm}}^{xx}=\widetilde{\chi}_{\mathrm{ee}}^{yy}$) \cite{Pfeiffer2013Huygens,Monticone2013,Pfeiffer2013Cascaded,Selvanayagam2013,Pfeiffer2013Millimeter,PfeifferNano2014,Wong2014,Epstein2016,Chen2018,Ataloglou2021}, under which backscattered waves due to the electric and magnetic responses perfectly cancel one another \cite{Love1976,Ziolkowski2010,Decker2015,Ziolkowski2022}, the grazing-angle Huygens' condition follows a similar mechanism at the grazing angle \cite{Shaham2024AdvOptMat}. The main difference that highlights the dominance of the latter is the increased sensitivity to its violation due to the singular wave immittance typical of near-grazing incidence, which results in high reflectivity unless adequately treated \cite{Shaham2024AdvOptMat}. Indeed, compliance with this extraordinary equilibrium has already been demonstrated to unlock all-angle transversely symmetric functionalities, such as all-angle transparency and all-angle PMC boundary conditions \cite{Shaham2024AdvOptMat,Shaham2025IEEETAP}. Therefore, the grazing-angle Huygens' condition of Eq.\ (\ref{Eq:GrazingHuygens}) is vital for asymmetric all-angle applications as well, including spatial differentiation.

Subject to Eq.\ (\ref{Eq:GrazingHuygens}),
%
%
the scattering at the grazing angle is no longer restricted to total reflection, but can be tuned according to $r(k_{x})\to\frac{r_{1}}{d_{1}}=\frac{4\widetilde{\chi}_{\mathrm{em}}^{yx}}{j\left[(\widetilde{\chi}_{\mathrm{em}}^{yx})^{2}-(\widetilde{\chi}_{\mathrm{mm}}^{xz})^{2}-4\right]}$ and $t(k_{x})\to\frac{j(t_0+t_1+t_2)}{d_{1}}=\frac{4-(\widetilde{\chi}_{\mathrm{em}}^{yx})^2+(\widetilde{\chi}_{\mathrm{mm}}^{xz})^2-4j\widetilde{\chi}_{\mathrm{mm}}^{xz}}{{(\widetilde{\chi}_{\mathrm{em}}^{yx})^{2}-(\widetilde{\chi}_{\mathrm{mm}}^{xz})^{2}-4}}$.
For a passive device without amplification, the presence of reflection comes at the cost of transmission. Furthermore, the desired transmission magnitude $|t(k_{x})|\propto|\widetilde{k}_{x}|$ increases with $|\widetilde{k}_{x}|$ and therefore should tend to its maximal value at the grazing angle $\widetilde{k}_{x}\to 1$. As so, in order to maximize the dynamic range of the processing through this passive device, we would like its transmittance at the grazing angle to be the largest possible without violation of power conservation, namely, $|t(k_{x})|\to 1$. Therefore, we eliminate reflection at that angle via
\begin{equation}
\label{Eq:NoBian}
    \widetilde{\chi}_{\mathrm{em}}^{yx}=-\widetilde{\chi}_{\mathrm{me}}^{xy}=0,
\end{equation}
i.e., vanishing omega bianisotropy. This sets our goal functionality to $t(k_{x})=-j\widetilde{k}_{x}$, soon to be achieved by enforcing additional relevant requirements on the susceptibilities.

Next, we notice that the transmission coefficient in Eq.\ (\ref{Eq:ScatCoeffs}) can be separated to its even and odd parts with respect to the transverse wavenumber $\widetilde{k}_{x}$: the former consists of the numerator terms with the coefficients $t_0$ and $t_{2}$, whereas the latter comprises the numerator term with the coefficient $t_{1}$. Since our goal $t(k_{x})=-j\widetilde{k}_{x}$ is purely odd, we eliminate the even coefficients by demanding $t_0=t_2=0$. Following Eq.\ (\ref{Eq:ScatCoeffs}), the first subrequirement, $t_0=0$, stands for vanishing transmission at the normal incidence, $t(k_{x}=0)=0$, and results in [Eq.\ (\ref{Eq:t_coeffs})]
\begin{equation}
\label{Eq:AntiHuygensNormal}
    \widetilde{\chi}_{\mathrm{ee}}^{yy}\widetilde{\chi}_{\mathrm{mm}}^{xx}=-4,
\end{equation}
as also reported in \cite{Holloway2005}. This can be viewed as an ``anti-Huygens' condition,'' under which the transmitted wave due to the electric MS response perfectly cancels that of the magnetic one for normal incidence, while the reflected waves interfere to provide unity reflectance, in a complementary analogy to the usual Huygens' condition of total transmission \cite{Pfeiffer2013Huygens,Monticone2013,Pfeiffer2013Cascaded,Selvanayagam2013,Pfeiffer2013Millimeter,PfeifferNano2014,Wong2014,Epstein2016,Chen2018,Ataloglou2021,Love1976,Ziolkowski2010,Decker2015,Ziolkowski2022}.

The second subcondition for suppressing the even quadratic part $\widetilde{k}_{x}^{2}$ in the numerator of the transmission coefficient, $t_{2}=0$, requires [Eq.\ (\ref{Eq:t_coeffs})]
\begin{equation}
\label{Eq:ZeroQuadTerm}
    (\widetilde{\chi}_{\mathrm{mm}}^{xz})^{2}=\widetilde{\chi}_{\mathrm{mm}}^{xx}\widetilde{\chi}_{\mathrm{mm}}^{zz}.
\end{equation}
This readily enforces $d_{3}=jt_{2}=0$ as well. Therefore, our requirements thus far lead to a
reduced form of the transmission coefficient, $t(k_{x})=\frac{jt_{1}\widetilde{k}_{x}}{d_{1}+d_{2}\widetilde{k}_{z}}=\frac{jt_{1}\widetilde{k}_{x}}{d_{1}+d_{2}\sqrt{1-\widetilde{k}_{x}^{2}}}$.
To obtain a pure linear dependence on $\widetilde{k}_{x}$ for perfect spatial differentiation, we also require $d_{2}=0$, i.e.,
\begin{equation}
\label{Eq:Vanishing_d2}
    \widetilde{\chi}_{\mathrm{mm}}^{xx}=\widetilde{\chi}_{\mathrm{mm}}^{zz}.
\end{equation}

Finally, combining the requirements of Eqs.\ (\ref{Eq:GrazingHuygens})--(\ref{Eq:Vanishing_d2})
yields a simple quadratic equation, whose solution is
\begin{equation}
\label{Eq:AllSolutions}
        \widetilde{\chi}_{\mathrm{mm}}^{xx}=\widetilde{\chi}_{\mathrm{mm}}^{zz}=-\widetilde{\chi}_{\mathrm{ee}}^{yy}=\pm2,
        \quad
        \begin{cases}
            \widetilde{\chi}_{\mathrm{mm}}^{xz}=-2\quad\Rightarrow\quad t(k_{x})=-j\widetilde{k}_{x}\\
            \widetilde{\chi}_{\mathrm{mm}}^{xz}=+2\quad\Rightarrow\quad t(k_{x})=+j\widetilde{k}_{x}.
        \end{cases}
\end{equation}
The top solution, $\widetilde{\chi}_{\mathrm{mm}}^{xz}=-2$, yields the desired perfect all-angle derivative operation $k_0^{-1}\partial_{x}=-j\widetilde{k}_{x}$, whereas the bottom solution, $\widetilde{\chi}_{\mathrm{mm}}^{xz}=+2$, yields its negative $-k_0^{-1}\partial_{x}=+j\widetilde{k}_{x}$. Note that the solution in Eq.\ (\ref{Eq:AllSolutions}) introduces an interesting degree of freedom, which is the possibility to choose the sign of the common value $\widetilde{\chi}_{\mathrm{mm}}^{xx}=\widetilde{\chi}_{\mathrm{mm}}^{zz}=-\widetilde{\chi}_{\mathrm{ee}}^{yy}$ at will ($2$ or $-2$). Both choices will satisfy the intricate balances required for perfect spatial differentiation in transmission; the only difference is the sign of the reflection coefficient, $r(k_{x})=j\mathrm{sgn}(\widetilde{\chi}_{\mathrm{mm}}^{xx})\widetilde{k}_{z}$. Note that the obtained condition is lossless, namely, the electric and magnetic (omega bianisotropic) responses assume purely real (imaginary) values \cite{Achouri2015,Achouri2018}; this can be easily verified by evaluating $|r(k_{x})|^{2}+|t(k_{x})|^{2}=|\widetilde{k}_{z}|^{2}+|\widetilde{k}_{x}|^{2}\equiv 1$ for all the angles of incidence $|\theta_0|<90^{\circ}$ [Eq.\ (\ref{Eq:NormalizedWavenumbers})], subject to Eq.\ (\ref{Eq:AllSolutions}). In practical realizations, small inevitable loss (e.g., conduction or dielectric losses in realistic materials) may inflict slight deviations from the ideal response; however, this deviation is generally expected to be small as long as low-loss materials are used and the meta-atom configuration is tuned sufficiently far from resonance, as inferred in \cite{Shaham2025IEEETAP}.

We have thus derived the rigorous closed-form relation (\ref{Eq:AllSolutions}) between the local MS susceptibility components for achieving perfect spatial differentiation with unity NA (accurately extending across the entire angular range, $|\theta_0|<90^{\circ}$). By this, not only have we perfected the NA performance of related past studies, e.g., \cite{Momeni2019,Abdolali2019,Momeni2021}, but we have also elucidated the distinct role of each underlying balance in the overall response and delineated a systematic algorithm for designing even more general asymmetric nonlocal MS functionalities in a likewise manner. In the next sections, we realize and validate this result by proposing, designing, and simulating two physical structures that support these balances: a conceptual meta-atom composed of a loaded loop and a loaded wire [Fig.\ \ref{Fig:Setup}(b), Sec.\ \ref{Subsec:MetaAtom}] and a more practical trilayered PCB MS of judiciously misaligned loaded-strip arrays [Fig.\ \ref{Fig:Setup}(c), Sec.\ {\ref{Subsec:PCB}}].

\section{Results and discussion}

\subsection{Local meta-atom design and validation}
\label{Subsec:MetaAtom}
For our first demonstration of realizing the perfect spatial derivative condition in Eq.\ (\ref{Eq:AllSolutions}), we propose the physical meta-atom configuration depicted in Fig.\ \ref{Fig:Setup}(b) for the frequency $f=20$ GHz (wavelength of $\lambda_0\approx 15$ mm in free space). Enclosed within periodic boundary conditions and subwavelength unit-cell dimensions of $L_{x}=L_{y}=3$ mm $\approx 0.2\lambda_0$, the meta-atom consists of a perfect-electric-conductor (PEC) loop loaded by two lumped capacitors of $C_{\mathrm{mm}}$ capacitance each, alongside a straight PEC line loaded by a lumped $C_{\mathrm{ee}}$ capacitance. The former (latter) is responsible for realizing the local magnetic (electric) responses of Eq.\ (\ref{Eq:AllSolutions}) in a well-controlled fashion of accurately tuning the capacitance values to meet the desired susceptibility values \cite{Tretyakov2003,Pfeiffer2013Huygens,Selvanayagam2013,Shaham2021,Shaham2024AdvOptMat}. To enable a nonvanishing tangential-normal magnetic component $\widetilde{\chi}_{\mathrm{mm}}^{xz}$ [specifically $\widetilde{\chi}_{\mathrm{mm}}^{xz}=-2$, Eq.\ (\ref{Eq:AllSolutions})], the loop is rotated by an angle $\alpha$ around the $y$-axis, whose value will soon be carefully fixed to meet our objectives; note that this rotation provides symmetry breaking along both the in-plane and out-of-plane axes, as generally required for asymmetric MS transmission, $t(-k_{x})\neq t(k_{x})$ \cite{Kwon2018,Momeni2019,Abdolali2019,Cordaro2019,Achouri2020,Shastri2023}.

\begin{figure}
    \includegraphics[width=\textwidth]{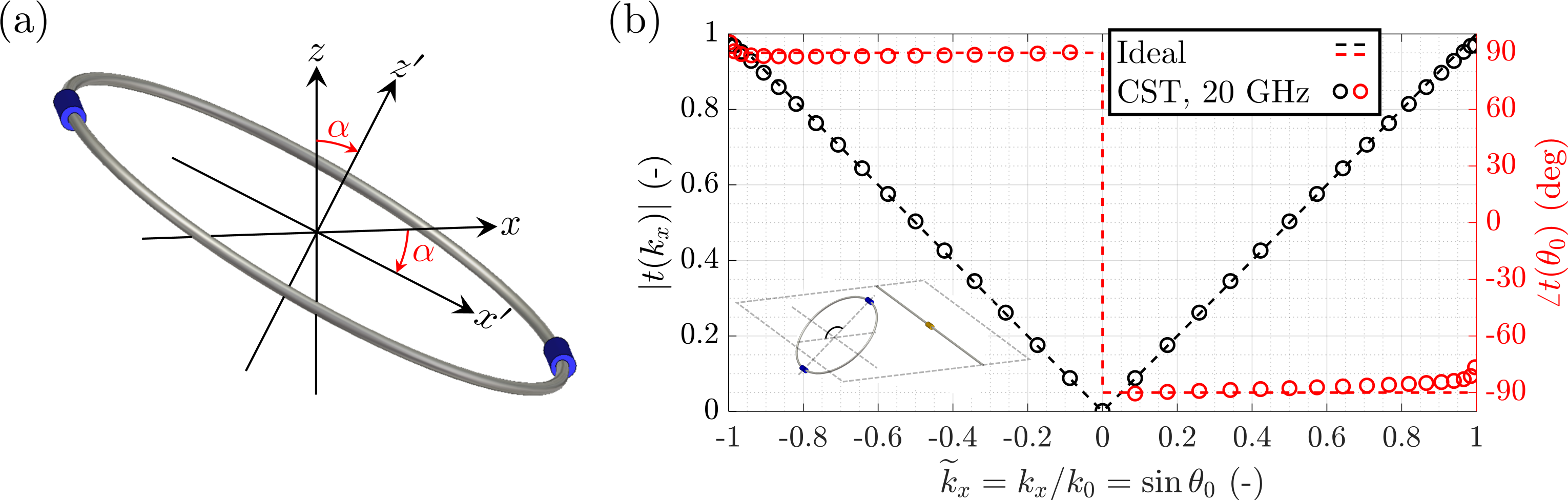}
    \caption{(a) Rotated loop geometry and coordinate system for manifesting anisotropy (Sec.\ \ref{Subsec:MetaAtom}). (b) Full-wave results (circle markers) for the 20-GHz transmission magnitude [$|t(k_{x)}|$, black, left abscissa] and phase [$\angle t(k_{x})$, red, right] versus tangential wavenumber (angle) through the rigorously tuned meta-atom in Fig.\ \ref{Fig:Setup}(b) in comparison to transmission through a perfect spatial differentiator $t(k_{x})=-j\widetilde{k}_{x}$ (dashed traces).
}
\label{Fig:MA_Results}
\end{figure}

We now proceed to tune the yet-unassigned parameters $\alpha$, $C_{\mathrm{mm}}$, and $C_{\mathrm{ee}}$. To this end, let us first understand the role of the rotation angle $\alpha$. Let us define an auxiliary coordinate system [Fig.\ \ref{Fig:MA_Results}(a)], wherein the normal to the plane of the loop is defined as the $z'$-axis, and where the $x'$-axis in the plane of the loop is determined such that $(x',y,z')$ is a right-handed system; specifically, for $\alpha=0$, the loop lies in the $xy$-plane, so that this auxiliary system coincides with the main coordinates, i.e., $(x',y,z')=(x,y,z)$. The loop is magnetically polarizable along the $z'$ direction (and no other direction), exhibiting an effective $\widetilde{\chi}_{\mathrm{mm}}^{z'z'}$ susceptibility value that depends on the $C_{\mathrm{mm}}$ capacitance value. This capacitance can be swept around its serial resonance with the effective self inductance of the loop, which remains constant once the outer radius, $r_{\mathrm{loop}}=1.35$ mm, wire diameter, $d_{\mathrm{wire}}=0.03$ mm, and frequency ($f=20$ GHz) are fixed \cite{Tretyakov2003,Shaham2021,Shaham2024AdvOptMat} [Fig.\ \ref{Fig:Setup}(b)]. This allows for a very wide range of $\widetilde{\chi}_{\mathrm{mm}}^{z'z'}$ to be realized \cite{Shaham2021,Shaham2024AdvOptMat}.

We now show that this $\widetilde{\chi}_{\mathrm{mm}}^{z'z'}$ susceptibility in the auxiliary coordinate system is transformed to exhibit $\widetilde{\chi}_\mathrm{mm}^{xx}$, $\widetilde{\chi}_\mathrm{mm}^{zz}$, and $\widetilde{\chi}_\mathrm{mm}^{xz}$ susceptibilities in the main coordinate system, which are related to one another through the angle $\alpha$. To this end, let us apply a $z$ directed average magnetic field $H_{z}^{\mathrm{av}}$ on the meta-atom. Its projection on the $z'$ axis is simply $H_{z'}^{\mathrm{av}}=\cos(\alpha)H_{z}^{\mathrm{av}}$. According to Eq.\ (\ref{Eq:Suscept}), a $z'$-directed surface magnetic polarization $M_{\mathrm{s}z'}=\chi_{\mathrm{mm}}^{z'z'}H_{z'}^{\mathrm{av}}=\chi_{\mathrm{mm}}^{z'z'}\cos(\alpha)H_{z}^{\mathrm{av}}$ is induced, which can be projected back onto the $x$ and $z$ directions via $M_{\mathrm{s}x}=\sin(\alpha)M_{\mathrm{s}z'}=\chi_{\mathrm{mm}}^{z'z'}\cos(\alpha)\sin(\alpha)H_{z}^{\mathrm{av}}$ and $M_{\mathrm{s}z}=\cos(\alpha)M_{\mathrm{s}z'}=\chi_{\mathrm{mm}}^{z'z'}\cos^{2}(\alpha)H_{z}^{\mathrm{av}}$. By the definition of Eq.\ (\ref{Eq:Suscept}), the last two relations yield the susceptibilities $\chi_{\mathrm{mm}}^{zz}=\cos^{2}(\alpha)\chi_{\mathrm{mm}}^{z'z'}$ and $\chi_{\mathrm{mm}}^{xz}=\cos(\alpha)\sin(\alpha)\chi_{\mathrm{mm}}^{z'z'}$ in the main coordinate system. Following a similar exercise of applying an $x$-directed average magnetic field $H_{x}^{\mathrm{av}}$ and employing similar projection operations yields the overall transformation
\begin{equation}
\label{Eq:RotatedSuscept}
    \begin{aligned}
        \widetilde{\chi}_{\mathrm{mm}}^{xx}&=\sin^{2}(\alpha)\widetilde{\chi}_{\mathrm{mm}}^{z'z'}, & \widetilde{\chi}_{\mathrm{mm}}^{zz}&=\cos^{2}(\alpha)\widetilde{\chi}_{\mathrm{mm}}^{z'z'}, & \widetilde{\chi}_{\mathrm{mm}}^{xz}&=\widetilde{\chi}_{\mathrm{mm}}^{zx}=\sin(\alpha)\cos(\alpha)\widetilde{\chi}_{\mathrm{mm}}^{z'z'}
    \end{aligned}
\end{equation}
(note that reciprocity is satisfied, as expected). In fact, this transformation is nothing more than a particular case of standard tensor rotation \cite{Lai2010},
\begin{equation}
\label{Eq:TensorRotation}
    \begin{bmatrix}
        \widetilde{\chi}_{\mathrm{mm}}^{xx} & \widetilde{\chi}_{\mathrm{mm}}^{xz}\\ \widetilde{\chi}_{\mathrm{mm}}^{zx} & \widetilde{\chi}_{\mathrm{mm}}^{zz}
    \end{bmatrix}
    =
    \begin{bmatrix}
        \cos\alpha & \sin\alpha \\ -\sin\alpha & \cos\alpha
    \end{bmatrix}
    \cdot
    \begin{bmatrix}
        \widetilde{\chi}_{\mathrm{mm}}^{x'x'} & \widetilde{\chi}_{\mathrm{mm}}^{x'z'}\\ \widetilde{\chi}_{\mathrm{mm}}^{z'x'} & \widetilde{\chi}_{\mathrm{mm}}^{z'z'}
    \end{bmatrix}
    \cdot
    \begin{bmatrix}
        \cos\alpha & \sin\alpha \\ -\sin\alpha & \cos\alpha
    \end{bmatrix}^{\mathrm{T}},
\end{equation}
with $\widetilde{\chi}_{\mathrm{mm}}^{x'x'}=\widetilde{\chi}_{\mathrm{mm}}^{x'z'}=\widetilde{\chi}_{\mathrm{mm}}^{z'x'}=0$ (as the loop is only polarizable along the $z'$-direction).

It follows from Eq.\ (\ref{Eq:RotatedSuscept}) that the susceptibility components exhibited by the rotated loop in Fig.\ \ref{Fig:MA_Results}(a) are constrained via
\begin{equation}
\label{Eq:SusceptRatios}
    \widetilde{\chi}_{\mathrm{mm}}^{xz}=\tan(\alpha)\widetilde{\chi}_{\mathrm{mm}}^{zz}=\cot(\alpha)\widetilde{\chi}_{\mathrm{mm}}^{xx}.
\end{equation}
In general, this constraint can be relaxed if an additional $x'$-polarizable loop is introduced as an additional degree of freedom [Eq.\ (\ref{Eq:TensorRotation})]. However, in our particular goal of a perfect all-angle spatially deriving MS, all the obtained specifications in Eq.\ (\ref{Eq:AllSolutions}) satisfy either $\widetilde{\chi}_{\mathrm{mm}}^{xz}=\widetilde{\chi}_{\mathrm{mm}}^{zz}=\widetilde{\chi}_{\mathrm{mm}}^{xx}$ or $\widetilde{\chi}_{\mathrm{mm}}^{xz}=-\widetilde{\chi}_{\mathrm{mm}}^{zz}=-\widetilde{\chi}_{\mathrm{mm}}^{xx}$; the former and latter are supported by the constraint in Eq.\ (\ref{Eq:SusceptRatios}) for $\alpha=45^{\circ}+180^{\circ}n$ and $\alpha=-45^{\circ}+180^{\circ}n$, respectively, where $n\in\mathbb{Z}$. For our meta-atom in Fig.\ \ref{Fig:Setup}(b), an angle of $\alpha=135^{\circ}$ is selected to realize a specific requirement from the set in Eq.\ (\ref{Eq:AllSolutions}),
\begin{equation}
\label{Eq:MetaAtomSolution}
    \widetilde{\chi}_{\mathrm{mm}}^{xx}=\widetilde{\chi}_{\mathrm{mm}}^{zz}=-\widetilde{\chi}_{\mathrm{mm}}^{xz}=-\widetilde{\chi}_{\mathrm{ee}}^{yy}=2.
\end{equation}

Once the angle $\alpha=-45^{\circ}$ is set, our next task is to tune the capacitance $C_{\mathrm{mm}}$ to realize the specific value $\widetilde{\chi}_{\mathrm{mm}}^{xx}=\widetilde{\chi}_{\mathrm{mm}}^{zz}=-\widetilde{\chi}_{\mathrm{mm}}^{xz}=2$. To this end, it is necessary to construct a look-up table (LUT) that relates between the capacitance value $C_{\mathrm{mm}}$ and the magnetic susceptibilities $\widetilde{\chi}_{\mathrm{mm}}^{xx}=\widetilde{\chi}_{\mathrm{mm}}^{zz}=-\widetilde{\chi}_{\mathrm{mm}}^{xz}$ associated with the loop \cite{Epstein2016,Shaham2021,Shaham2024AdvOptMat}. Note that constructing a LUT only for the tangential component $\widetilde{\chi}_{\mathrm{mm}}^{xx}$ suffices, as the rest are readily determined from it via Eq.\ (\ref{Eq:SusceptRatios}).

In general, to perform an accurate such characterization, it is typically necessary to model the meta-atom in a full-wave solver and simulate for its scattering properties under periodic boundary conditions \cite{Epstein2016}; the LUT is established by repeating this procedure for different values of the meta-atom parameters under inspection (in our case $C_{\mathrm{mm}}$). Specifically, if the meta-atom is guaranteed to exhibit only the $\widetilde{\chi}_{\mathrm{ee}}^{yy}$, $\widetilde{\chi}_{\mathrm{mm}}^{xx}$, and $\widetilde{\chi}_{\mathrm{mm}}^{zz}$ responses (and no other components, in our case $\widetilde{\chi}_{\mathrm{mm}}^{xz}=\widetilde{\chi}_{\mathrm{em}}^{yx}=0$), then these susceptibilities can be extracted from the simulated plane-wave scattering parameters (specular reflection and direct transmission) at normal incidence $\theta_0=0$ and at one additional oblique angle $\theta_0\neq 0$ \cite{Holloway2009,Zaluski2016,Shaham2021,Shaham2024AdvOptMat,Shaham2025IEEETAP} (see Appendix \ref{Subsec:Char}). Herein, however, the tangential-normal component $\widetilde{\chi}_{\mathrm{mm}}^{xz}$ is expected to be nonvanishing, such that, in principle, the characterization methodology must be extended to support this scenario as well. Nevertheless, since, in this particular case, we are interested in characterizing only the tangential component $\widetilde{\chi}_{\mathrm{mm}}^{xx}$, it is possible to simulate for the scattering parameters only at normal incidence $\theta_0=0$ and follow the formulas of the standard characterization scheme \cite{Holloway2009,Zaluski2016,Shaham2021,Shaham2024AdvOptMat,Shaham2025IEEETAP} (Appendix \ref{Subsec:Char}).

Therefore, we model the meta-atom in Fig.\ \ref{Fig:Setup}(b) in the commercial full-wave solver ``CST Studio Suite'' (CST) and simulate for the specular reflection $r(0)$ and direct transmission $t(0)$ due to a normally incident plane wave of $f=20$ GHz under periodic boundary conditions. We then substitute the results in Eq.\ (\ref{Eq:TangentialChar}), Appendix \ref{Subsec:Char}, to extract the associated $\widetilde{\chi}_{\mathrm{mm}}^{xx}$ susceptibility versus the capacitance $C_{\mathrm{mm}}$ and establish the desired LUT. Note that the magnetic susceptibilities, in particular $\widetilde{\chi}_{\mathrm{mm}}^{xx}$, are virtually independent of the load capacitance $C_{\mathrm{ee}}$ of the electrically polarizable line. The reason is that each loop in this periodic structure is centered between two adjacent lines, such that their contributions to the magnetic flux through it perfectly cancel each other. From the LUT, we find that $C_{\mathrm{mm}}=35.255$ fF yields $\widetilde{\chi}_{\mathrm{mm}}^{xx}\approx2$, as desired in Eq.\ (\ref{Eq:MetaAtomSolution}), and fix it.

Similarly, we may independently tune the remaining $\widetilde{\chi}_{\mathrm{ee}}^{yy}$ component through the value of $C_{\mathrm{ee}}$. Specifically, we establish another LUT of $\widetilde{\chi}_{\mathrm{ee}}^{yy}$ versus $C_{\mathrm{ee}}$ by substituting the simulated scattering parameters of the meta-atom for normal incidence in Eq.\ (\ref{Eq:TangentialChar}), Appendix \ref{Subsec:Char}, for each $C_{\mathrm{ee}}$ value under inspection. We find that $C_{\mathrm{ee}}=90$ fF yields $\widetilde{\chi}_{\mathrm{ee}}^{yy}\approx -2$, as desired, and fix it as well. We thus finalize the meta-atom design, which, according to the above characterization method, is expected to accurately implement the specifications in Eq.\ (\ref{Eq:MetaAtomSolution}) and realize perfect all-angle spatial differentiation $t(k_{x})=-j\widetilde{k}_{x}$.

To demonstrate the validity and accuracy of our methodology, we inspect the finalized meta-atom in CST and plot the full-wave magnitude $|t(k_{x})|$ (black plots, left abscissa) and phase $\angle t(k_{x})$ (red plots, right abscissa) of the transmission coefficient versus the tangential wavenumber $\widetilde{k}_{x}=\sin\theta_0$ in Fig.\ \ref{Fig:MA_Results}(b); the angle $\theta_0$ is swept from $-85^{\circ}$ to $85^{\circ}$ in steps of $5^{\circ}$. We compare these results to their corresponding ideal values, $|t(k_{x})|=|\widetilde{k}_{x}|$ and $\angle t(k_{x})=\angle (-j\widetilde{k}_{x})=-90^{\circ}\mathrm{sgn}(\widetilde{k}_{x})$ (dashed traces). Indeed, we observe excellent agreement in both magnitude and phase for all angles. Specifically, near-unity magnitudes (and excellent phase accuracy) are achieved close to the grazing angles ($\widetilde{k}_{x}\approx\pm 1$), verifying that this MS spatial differentiator is of $\mathrm{NA}\approx 1$. 

Hence, we have thus far established that perfect spatial differentiation with near-unity NA can not only be achieved in closed form, but can also be rigorously realized using electromagnetically justified meta-atoms carefully engineered to satisfy this criterion. From a broader perspective, this demonstration validates a key functionality supported by the general framework of asymmetric transmission engineering presented in Sec.\ \ref{Subsec:GeneralASymmetric} and delineates a systematic realization scheme. Having said that, the meta-atom in Fig.\ \ref{Fig:Setup}(b) is difficult to realize in practice as it is generally not supported by widely available fabrication technologies. Nevertheless, the crucial observations and insights gained by this meta-atom demonstration will play an important role in our next design of a more practical PCB version of this MS.

\subsection{Nonlocal PCB MS design and validation}
\label{Subsec:PCB}
Having formulated the closed-form electromagnetic conditions for perfect spatial differentiation and validated them at the meta-atom level, we next employ the principles developed above in designing a more practical MS compatible with PCB technology. For quite a long time, PCBs have offered an appealing platform for implementing MSs at the radio and microwave frequencies \cite{Glybovski2016,Chen2016,Li2018,Quevedo-Teruel2019}. In particular, PCB Huygens' and omega-bianisotropic MSs \cite{Pfeiffer2013Cascaded,Pfeiffer2013Millimeter,Epstein2016,Epstein2016OBMS,Epstein2016PRL,Chen2018,Ataloglou2021} have drawn considerable attention for their utility in transmissive beam manipulation.

The basic PCB configuration of a TE-polarized Huygens' MS is realized via three layers of printed loaded-strip arrays separated by thin dielectric spacers \cite{Pfeiffer2013Cascaded,Pfeiffer2013Millimeter,Epstein2016}. When the arrays are sufficiently dense, such that higher-order diffracted modes decay rapidly, each of the layers can be represented by a homogeneous tangential electric response independent of the other layers \cite{Tretyakov2003}. The symmetric mode supported by the thin cascade, in which the surface currents in all the layers flow in the same direction, manifests effective electric response (similarly to a single current sheet); at the same time, the antisymmetric mode, where the surface currents in the outer layers are perfectly opposed to each other, can be viewed as an effective loop that emulates the tangential magnetic counterpart \cite{Monticone2013,Pfeiffer2013Cascaded,Pfeiffer2013Millimeter,Epstein2016,Chen2018,Ataloglou2021}.

For many years, this decomposition has served as the main principle behind realizing PCB Huygens' MSs, specifically by balancing these responses at normal incidence to nullify reflection. If the longitudinal symmetry is broken (the properties of the two outer layers differ from one another), then tangential omega bianisotropy arises, which can be leveraged for wide-angle and nonlocal-power-redistributing transformations \cite{Epstein2016OBMS,Epstein2016PRL}. However, as we have recently shown in \cite{Shaham2024AdvOptMat,Shaham2025IEEETAP}, such admittance-sheet cascades exhibit not only the previously reported effective tangential responses $\chi_{\mathrm{mm}}^{yy}$, $\chi_{\mathrm{mm}}^{xx}$, and $\chi_{\mathrm{me}}^{xy}=-\chi_{\mathrm{em}}^{yx}$, but also the effective normal component $\chi_{\mathrm{mm}}^{zz}$ that, together with the rest, accurately captures the  scattering behavior of the PCB MS in closed form for all the angles at once ($|\theta_0|<90^{\circ}$). This extended feature, enabled by nonlocal mechanisms of multiple reflections between the layers \cite{Shastri2023,Shaham2024AdvOptMat,Shaham2025IEEETAP}, has unlocked new powerful possibilities to defy the seemingly persistent reflection near the grazing angles [via the grazing-angle Huygens' condition, Eq.\ (\ref{Eq:GrazingHuygens})] and thus realize all-angle functionalities on demand. 

However, as long as the homogenization approximation of the layers is valid, this scheme is limited to support only functionalities of symmetric transmission $t(-k_{x})=t(k_{x})$, since the scattering off tangentially polarizable impedance sheets is itself $k_{x}$-symmetric. To allow for asymmetric transmission in the MS regime, where only the fundamental diffraction order propagates, not only must the strip arrays be misaligned with respect to one another, but they must be closely spaced to allow significant interlayer interactions through the higher-order evanescent nearfields \cite{Molero2017,Xu2021}. By this violation of the layer homogenization, we may drastically enrich the overall spatial dispersion of the composite and realize asymmetric transmission. While the modeling of such PCB loaded-strip arrays---complex as they may be---is quite well-established \cite{Wait1955,Tretyakov2003}, mostly for metagrating purposes, e.g., \cite{Rabinovich2017,Molero2017,Rabinovich2018Analytical,Rabinovich2018Arbitrary,Xu2021,Yashno2023}, their design is often limited to control a single or a finite number of incidence scenarios (often referred to as multichannel metagratings). Hence, the path for utilizing these models to accomplish all-angle operation is currently veiled.

Inspired by the ability of thin loaded-strip PCB stacks to accurately emulate symmetric tangential and normal MS susceptibilities for all angles \cite{Shaham2025IEEETAP}, we propose to modify the configuration in  \cite{Shaham2025IEEETAP} to introduce and control also the asymmetric $\widetilde{\chi}_{\mathrm{mm}}^{xz}=\widetilde{\chi}_{\mathrm{mm}}^{zx}$ component by judiciously misaligning the arrays and carefully tuning their constituent parameters. To demonstrate both the design procedure and its viability, we propose the PCB unit-cell configuration in Fig.\ \ref{Fig:Setup}(c) for realizing a highly accurate high-NA all-angle (TE) spatial differentiator. The unit-cell of $L_{x}\times L_{y}$ area consists of three layers---bottom ($z=-d$), middle ($z=0$), and top ($z=d$)---each of which accommodates one $y$-directed $18$-$\mu$m-thick printed copper strip of width $w$ located at $x=x_{\mathrm{bot}}$, $x=0$, and $x=x_{\mathrm{top}}$ and loaded by printed capacitors of width $W_{\mathrm{bot}}$, $W_{\mathrm{mid}}$, and $W_{\mathrm{top}}$, respectively. The layers are separated by dielectric substrates of relative permittivity $\epsilon_{\mathrm{r}}$ and a deeply subwavelength thickness $d$ each, such that the overall thickness of the device is $2d$.

The composite is excited from below ($z<-d$) by a TE $y$-polarized plane wave $E_{y}^{\mathrm{inc}}(\vec{r})=E_0e^{-j[k_{x}x+k_{z}(z+d)]}$, where $E_0$ is the amplitude at $x=0$ on the reference plane $z=-d$, and $k_{x}=k_0\sin\theta_0$ and $k_{z}=k_0\cos\theta_0$ are the wavevector components determined by the angle of incidence $\theta_0$, as before. The electromagnetic scattering in this problem can be accurately obtained via standard Floquet-Bloch (FB) analysis, e.g., \cite{Rabinovich2018Analytical,Rabinovich2018Arbitrary,Xu2021,Yashno2023}, as follows. Since the $y$-periodicity is set deeply subwavelength, $L_{y}=1.8$ mm $\approx 0.12\lambda_0$, the impedance of the capacitive load can be homogeneously distributed per-unit-length along the strip, such that the scattered fields remain $y$-polarized, and $y$-independent (macroscopically). Furthermore, due to the $L_{x}$ periodicity along $x$, the scattered fields everywhere can be expressed via a FB series of plane waves. Each such FB plane wave is associated with a tangential and a normal wavenumber, $k_{x,n}=k_{x}+2\pi n/L_{x}$ and
\begin{equation}
\label{Eq:kzn}
    k_{z,n}=
    \begin{cases}
        \sqrt{k_{0}^{2}-\epsilon_{r}k_{x,n}^{2}}, & |z|<d\\
        \sqrt{k_{0}^{2}-k_{x,n}^{2}}, & |z|>d.
    \end{cases}
\end{equation}

The $x$-periodicity is set smaller than half the wavelength in each medium, namely, $L_{x}<\frac{\lambda_0}{2},\frac{\lambda_0}{2\sqrt{\epsilon_{\mathrm{r}}}}$; hence, only the fundamental ($n=0$) order propagates and the higher-order harmonics ($n\neq 0$) are evanescent for all the angles of incidence. In free space, the evanescent fields decay rapidly away from the MS facets, such that the scattered fundamental harmonics ($n=0$) dominate and can be expressed as specular reflection, $E_{y}^{\mathrm{ref}}(\vec{r})=r(k_{x})E_0e^{-j[k_{x}x-k_{z}(z+d)]}$, and direct transmission, $E_{y}^{\mathrm{tran}}(\vec{r})=t(k_{x})E_0e^{-j[k_{x}x+k_{z}(z-d)]}$, whose amplitudes are defined at $x=0$ on the reference planes $z=-d$ and $z=d$, respectively. The geometric and material parameters of the PCB determine the values of $t(k_{x})$ and $r(k_{x})$, which can be calculated by the FB analysis in \cite{Rabinovich2018Arbitrary}.

Our task now is to set these parameters to aspire $t(k_{x})=-j\widetilde{k}_{x}$, as in Sections \ref{Subsec:SpatialDiff} and \ref{Subsec:MetaAtom}. To this end, we aim at effectively realizing one set of the perfect all-angle spatial differentiation conditions in Eq.\ (\ref{Eq:AllSolutions}) through this structure (to the best accuracy possible). Note that a tacit assumption is made that the PCB structure in Fig.\ \ref{Fig:Setup}(c) can indeed exhibit local effective $\widetilde{\chi}_{\mathrm{ee}}^{yy}$, $\widetilde{\chi}_{\mathrm{mm}}^{xx}$, $\widetilde{\chi}_{\mathrm{mm}}^{zz}$, and $\widetilde{\chi}_{\mathrm{mm}}^{xz}=\widetilde{\chi}_{\mathrm{mm}}^{zx}$ components while suppressing other undesired response types for all angles. This assumption is based on its validity for the particular case of aligned strips ($x_{\mathrm{top}}=x_{\mathrm{bot}}=0$), as rigorously derived in \cite{Shaham2025IEEETAP}, and will evidently yield very reliable predictions in what follows. Note also that this assumption is specific to the unique scenario of misaligned strip arrays and may not necessarily hold for other, more general geometries.

Let us first fix the frequency to $f=20$ GHz ($\lambda_0\approx 15$ mm), the substrate parameters to those of a commercially available Rogers RO3003 laminate of relative permittivity $\epsilon_{\mathrm{r}}=3$ and thickness $d=20$ mil $=0.508$ mm $\approx 0.034\lambda_0\approx 0.0587\frac{\lambda_0}{\sqrt{\epsilon_{\mathrm{r}}}}$, and the strip width to $w=5$ mil $=0.127$ mm (supported by standard PCB fabrication technology). Next, to ensure suppression of the omega-bianisotropic component ($\widetilde{\chi}_{\mathrm{me}}^{xy}=\widetilde{\chi}_{\mathrm{em}}^{yx}=0$), we invoke that the MS scattering properties must remain identical when the illumination from below the MS ($z<0$) at angle $\theta_0$ is replaced by another from above ($z>0$) with angle $\theta_0+180^{\circ}$ \cite{Achouri2020}. This will be guaranteed if we constrain equal and opposite lateral shifts of the top and bottom strips, $x_{\mathrm{top}}=-x_{\mathrm{bot}}=\Delta x$, and equal load capacitors, $W_{\mathrm{top}}=W_{\mathrm{bot}}=W$; these common values will now serve as degrees of freedom to be tuned, along with the periodicity $L_{x}$ and the capacitor width of the center layer, $W_{\mathrm{mid}}$.

Note the resemblance between the PCB stack in Fig.\ \ref{Fig:Setup}(c) and the meta-atom of rotated loaded loop and centered loaded line in Fig.\ \ref{Fig:Setup}(b). At the intuitive level, the aforementioned antisymmetric mode \cite{Monticone2013,Pfeiffer2013Cascaded,Pfeiffer2013Millimeter,Epstein2016,Chen2018,Ataloglou2021} (where opposing currents are induced through the top and bottom strips) now yields a rotated effective loop. Hence, tuning the geometric dimensions $L_{x}$ and $\Delta x$ can be thought of as analogous, in a sense, to tuning the rotation angle $\alpha$ of the meta-atom loop, while tuning the PCB capacitor widths is analogous to that in the meta-atom design. In light of this analogy and the tuning procedure in Sec.\ \ref{Subsec:MetaAtom}, our first task is to find $L_{x}$ and $\Delta x$ values that would ensure adequate balances between the effective susceptibilities [in the spirit of Eq.\ (\ref{Eq:SusceptRatios})] and facilitate approaching one of the perfect-derivative conditions in Eq.\ (\ref{Eq:AllSolutions}), specifically Eq.\ (\ref{Eq:MetaAtomSolution}).

To this end, we need to be able to assess the effective $\widetilde{\chi}_{\mathrm{ee}}^{yy}$, $\widetilde{\chi}_{\mathrm{mm}}^{xx}$, $\widetilde{\chi}_{\mathrm{mm}}^{zz}$, and $\widetilde{\chi}_{\mathrm{mm}}^{xz}=\widetilde{\chi}_{\mathrm{mm}}^{zx}$ susceptibilities associated with a candidate PCB setup. We therefore extend the characterization scheme in \cite{Holloway2009,Zaluski2016,Shaham2021,Shaham2024AdvOptMat,Shaham2025IEEETAP}, which supports only the three former symmetric components, to also support the latter antisymmetric susceptibility (Appendix \ref{Subsec:Char}). In this extension, the specular reflection $r(k_{x})$ and direct transmission $t(k_{x})$ of the fundamental FB harmonic must be retrieved for three angles of incidence: normal incidence ($\theta_0=0$), another oblique angle ($\theta_0$, herein $\theta_0=30^{\circ}$ is selected), and its negative ($-\theta_0$). These data can then substituted in Eqs.\ (\ref{Eq:TangentialChar}), (\ref{Eq:Char_mm_zz}), and (\ref{Eq:LinearRelationsPositive}) to extract the four values $\widetilde{\chi}_{\mathrm{ee}}^{yy}$, $\widetilde{\chi}_{\mathrm{mm}}^{xx}$, $\widetilde{\chi}_{\mathrm{mm}}^{zz}$, and $\widetilde{\chi}_{\mathrm{mm}}^{xz}=\widetilde{\chi}_{\mathrm{mm}}^{zx}$ (see Appendix \ref{Subsec:Char}).

For each candidate set of parameters in Fig.\ \ref{Fig:Setup}(c), we evaluate these scattering data via the FB analysis in \cite{Rabinovich2018Analytical,Rabinovich2018Arbitrary,Yashno2023} and then substitute them in the aforementioned characterization equations. Next, aiming at the requirement of Eq.\ (\ref{Eq:MetaAtomSolution}), we use the \texttt{fmincon} optimization function in \MATLAB to minimize the least-squared-error cost function $(\widetilde{\chi}_{\mathrm{ee}}^{yy}+2)^{2}+(\widetilde{\chi}_{\mathrm{mm}}^{xx}-2)^{2}+(\widetilde{\chi}_{\mathrm{mm}}^{zz}-2)^{2}+(\widetilde{\chi}_{\mathrm{mm}}^{xz}+2)^{2}$. The geometric parameters are swept in the range $0.1\lambda_0\leq L_{x}\leq 0.45\lambda_0$ and $|\Delta x|\leq L_{x}/4$, whereas, for simplicity, the load parameters $W$ and $W_{\mathrm{mid}}$ are rather described through the total equivalent impedance of the strips, $Z_{\mathrm{top}}=Z_{\mathrm{bot}}=jX$ and $Z_{\mathrm{mid}}=jX_{\mathrm{mid}}$, due to their self inductance in series with their load capacitance. No restrictions are imposed on the (real) values of $X$ and $X_{\mathrm{mid}}$ during sweeping. Fifty random initial guesses are taken to ensure convergence to the global minimum. The optimization yields the parameter values $L_{x}\approx 7.51 d\approx 3.815$ mm $\approx 0.254 \lambda_0$, $\Delta x\approx 0.77 d\approx 0.1025L_{x}\approx 0.391$ mm, $X \approx -315.79\eta_0$, and $X_{\mathrm{mid}}\approx -214.96 \eta_0$, which attain the effective susceptibilities $\widetilde{\chi}_{\mathrm{ee}}^{yy}\approx-1.991$, $\widetilde{\chi}_{\mathrm{mm}}^{xx}\approx 2.1456$, $\widetilde{\chi}_{\mathrm{mm}}^{zz}\approx2.094$, and $\widetilde{\chi}_{\mathrm{mm}}^{xz}=\widetilde{\chi}_{\mathrm{mm}}^{zx}\approx -1.76$, plausibly close to our goal in Eq.\ (\ref{Eq:MetaAtomSolution}).

\begin{figure}
    \includegraphics[width=\textwidth]{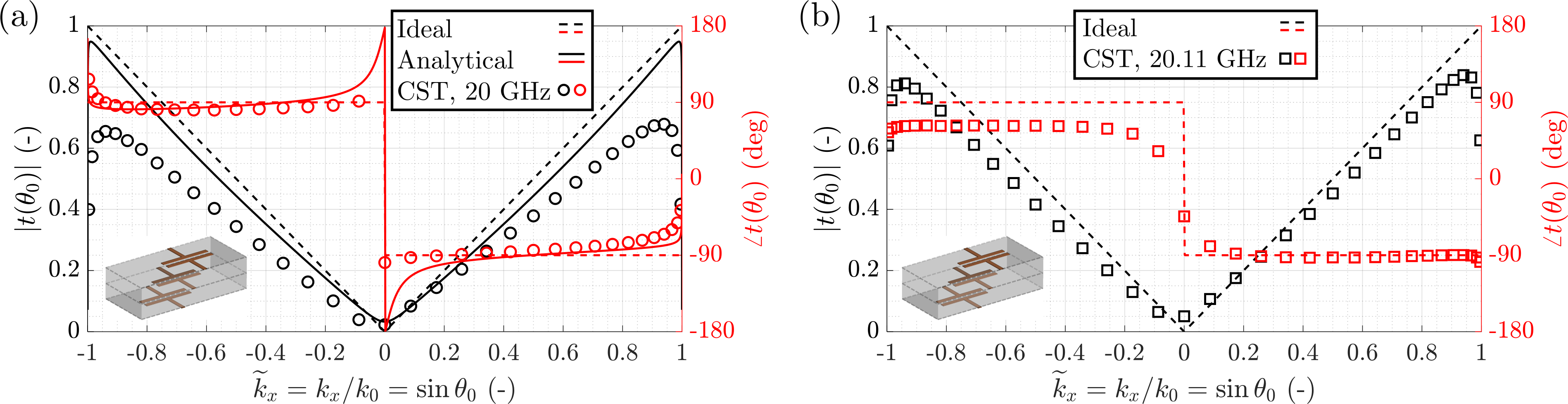}
    \caption{(a) Analytical (solid traces, evaluated via an FB analysis \cite{Rabinovich2018Analytical,Rabinovich2018Arbitrary,Yashno2023}) and full-wave (circle markers) results for the 20-GHz transmission magnitude [$|t(k_{x)}|$, black, left abscissa] and phase [$\angle t(k_{x})$, red, right] versus tangential wavenumber (angle) for the rigorously tuned PCB in Fig.\ \ref{Fig:Setup}(c) in comparison to transmission through a perfect spatial differentiator $t(k_{x})=-j\widetilde{k}_{x}$ (dashed traces) (b) Full-wave results (circle markers) for $f=20.11$ GHz compared to ideal values (dashed traces).
}
\label{Fig:PCB_Results}
\end{figure}

Figure \ref{Fig:PCB_Results}(a) compares the resulting analytical (solid traces, FB analysis as in \cite{Rabinovich2018Analytical,Rabinovich2018Arbitrary,Yashno2023}) transmission magnitude [$|t(k_{x})|$, black plots, left abscissa] and phase [$\angle t(k_{x})$, red plots, right abscissa] versus $\widetilde{k}_{x}=\sin\theta_0$ to the ideal response [$t(k_{x})=-j\widetilde{k}_{x}$, dashed traces] for $f=20$ GHz. Indeed, very good performance is observed in both. In particular, the analytical FB transmission magnitude is extremely close to linear, peaking at approximately $0.987$ (near-unity) for $\widetilde{k}_{x}\approx\pm 0.948$ and dropping to approximately $0.033$ (near-zero) at normal incidence, as desired. For $|\widetilde{k}_{x}|>0.11$, where $|t(k_{x})|>0.1$, the analytical FB transmission phase is accurate up to maximal and typical deviations of $\pm 20^{\circ}$ and $\pm 10^{\circ}$, respectively. Despite this minor deviation due to effective susceptibility imperfections, the respectably accurate and high-NA performance validates the potential to achieve nearly perfect spatial differentiation with such PCB MSs.

Finally, to further validate our results with realistic printed capacitors and subject to inevitable copper and dielectric loss, we first fix the above optimal dimensions, $L_{x}=3.815$ mm and $\Delta x=0.391$ mm; we then characterize the MS at normal incidence in CST via Eq.\ (\ref{Eq:TangentialChar}), while $W$ and $W_{\mathrm{mid}}$ are swept, similarly to the meta-atom design procedure in Sec.\ \ref{Subsec:MetaAtom}. Specifically, since the effective value $\widetilde{\chi}_{\mathrm{mm}}^{xx}$ depends only on the common value $W$ \cite{Shaham2025IEEETAP}, we first tune it to $W=1.62$ mm ($\widetilde{\chi}_{\mathrm{mm}}^{xx}\approx 1.98$); then, we tune $W_{\mathrm{mid}}=1.52$ mm ($\widetilde{\chi}_{\mathrm{ee}}^{yy}\approx-2.016$).

We inspect the full-wave transmission of this finalized realistic PCB design at $f=20$ GHz (circle markers) in Fig.\ \ref{Fig:PCB_Results}(a), comparing them to their ideal (dashed traces) and analytically calculated (solid traces) counterparts. The full-wave transmission phase follows its analytically calculated counterpart, featuring even better agreement with ideal values near the normal incidence. The transmission magnitude retains very good linearity for $|\widetilde{k}_{x}|\leq 0.94$, peaking, however, at a reduced value of approximately $0.66$. This minor deviation from the analytical predictions may be mainly attributed to inherent inaccuracies in the analytical models used herein \cite{Rabinovich2018Analytical,Rabinovich2018Arbitrary,Shaham2025IEEETAP}. Nevertheless, the high linearity of the magnitude response with respect to $|\widetilde{k}_{x}|$ and plausibly accurate phase response already make this device a well-performing spatial differentiator with near-unity NA, albeit with approximately $66\%$ of the ideal dynamic range.

As presented in Fig.\ \ref{Fig:PCB_Results}(b), the dynamic range is significantly improved at the very slightly shifted frequency $f=20.11$ GHz ($<0.6\%$ deviation), retaining excellent linearity with respect to $|\widetilde{k}_{x}|$ and reaching approximately $0.82$ transmission magnitude at $\widetilde{k}_{x}\approx\pm 0.94$. At this frequency, the phase stabilizes tightly around $-92^{\circ}$ (extremely close to the desired $-90^{\circ}$ value) for positive transverse wavenumbers $\widetilde{k}_{x}$, and around $+62^{\circ}$ (deviating from the desired $+90^{\circ}$ value) for negative $\widetilde{k}_{x}$ values. This phase deviation may be acceptable for many applications. For instance, the well-known Fraunhofer distance that separates between the nearfield and farfield regions of antennas and optical apertures is customarily based on maximal phase deviation of $22.5^{\circ}$ ($\pi/8$ radians) in the spectral components, often yielding a good approximation of the radiation pattern or image \cite{Balanis2016,Goodman2017}. Overall, our results in this demonstration validate that highly accurate TE spatial differentiation with nearly unity NA can be achieved based on the universal susceptibility criterion of Eq.\ (\ref{Eq:MetaAtomSolution}) via the widely accessible platform of multilayered PCBs.

\section{Conclusion}
To conclude, we have established a rigorous and systematic framework for realizing asymmetric all-angle transmission in MSs, with perfect first-order spatial differentiation as a representative and practically relevant example. By extending the recently introduced all-angle MS formalism in \cite{Shaham2025IEEETAP} to asymmetric scenarios, especially the viability of the grazing-angle Huygens' condition \cite{Shaham2024AdvOptMat,Shaham2025IEEETAP}, we have derived closed-form susceptibility conditions that guarantee an exact target response over the entire propagating spectrum. This analytical solution provides both physical insights and a well-defined design scheme, eliminating the need for ad-hoc optimization procedures.

Validating this theoretical formulation, we have demonstrated that the derived conditions are supported by physically meaningful meta-atom implementations, ranging from conceptual local inclusions to realistic nonlocal multilayered PCB architectures. In contrast to earlier demonstrations relying on intricate or fabrication-challenging geometries, the proposed structures are directly connected to the underlying susceptibility requirements and can be systematically engineered using well-understood electromagnetic mechanisms. While spatial differentiation is used here as a canonical asymmetric functionality, the presented methodology is general and modular, and is therefore readily extendable to a broader class of asymmetric nonlocal operations, including higher-order processing and other wave-based analog functionalities. Such high-NA (unity-NA, in principle) performance enables high resolution in wave and image processing beyond the paraxial regime. In the future, it may be possible to extend this methodology to support also the transverse-magnetic (TM)  polarization and even translate it to the optical domain, e.g., by preserving the overall geometric principles while transitioning to low-loss dielectric platforms. It could also be possible to consider introducing more degrees of freedom to manage loss and bandwidth aspects, e.g., more loaded strips or layers, as in \cite{Pfeiffer2014Bian,Sanchez2018}. Collectively, these prospects position the present approach as a versatile foundation for next-generation high-NA nonlocal MSs and asymmetric wave-processing devices.

\appendix
\section{Appendix: Extraction of associated surface susceptibilities}

\label{Subsec:Char}
In this section, we extend the susceptibility characterization method in \cite{Holloway2009,Zaluski2016,Shaham2021,Shaham2024AdvOptMat,Shaham2025IEEETAP}---which supports only the extraction of the $\chi_{\mathrm{ee}}^{yy}$, $\chi_{\mathrm{mm}}^{xx}$, and $\chi_{\mathrm{mm}}^{zz}$ components and assumes that the others vanish---to accommodate our needs in this work. Focusing on non-bianisotropic applications ($\overline{\overline{\chi}}_{\mathrm{me}}=\overline{\overline{\chi}}_{\mathrm{em}}=0$), e.g., spatial differentiation (Sections \ref{Subsec:SpatialDiff}--\ref{Subsec:PCB}), we assume that the device under test features these three former susceptibilities in addition to nonvanishing $\chi_{\mathrm{mm}}^{xz}=\chi_{\mathrm{mm}}^{zx}$ components and that all other response types vanish.

Similarly to \cite{Holloway2009,Zaluski2016}, we first retrieve the full-wave specular reflection $r(0)$ and direct transmission $t(0)$ coefficients for a normally incident plane wave ($\theta_0=0$), i.e., for $\widetilde{k}_{x}=\sin\theta_0=0$ and $\widetilde{k}_{z}=\cos\theta_0=1$. According to Eqs.\ (\ref{Eq:ScatCoeffs})--(\ref{Eq:t_coeffs}) these values are related only to the tangential susceptibilities via
\begin{equation}
\label{Eq:ScatNormal}
    \begin{aligned}
        r(0)&=\frac{r_0+r_2}{d_0+d_1+d_2+d_3}=\frac{2(\widetilde{\chi}_{\mathrm{mm}}^{xx}-\widetilde{\chi}_{\mathrm{ee}}^{yy})}{2(\widetilde{\chi}_{\mathrm{mm}}^{xx}+\widetilde{\chi}_{\mathrm{ee}}^{yy})+j(\widetilde{\chi}_{\mathrm{ee}}^{yy}\widetilde{\chi}_{\mathrm{mm}}^{xx}-4)}, \\
        t(0)&=\frac{jt_0}{d_0+d_1+d_2+d_3}=\frac{-j({\chi}_{\mathrm{ee}}^{yy}\widetilde{\chi}_{\mathrm{mm}}^{xx}+4)}{2(\widetilde{\chi}_{\mathrm{mm}}^{xx}+\widetilde{\chi}_{\mathrm{ee}}^{yy})+j(\widetilde{\chi}_{\mathrm{ee}}^{yy}\widetilde{\chi}_{\mathrm{mm}}^{xx}-4)}.
    \end{aligned}
\end{equation}
We therefore obtain a set of two equations relating between the two unknown tangential susceptibilities. Note that these relations are completely indifferent to the values of the normal susceptibilities (since normal incidence has zero normal component of the magnetic field); as so, the characterization of these responses reduces exactly to the one obtained in \cite{Holloway2009,Zaluski2016},
\begin{equation}
\label{Eq:TangentialChar}
    \begin{aligned}
    \widetilde{\chi}_{\mathrm{ee}}^{yy}&=2j\cdot\frac{t(0)+r(0)-1}{t(0)+r(0)+1}, & \widetilde{\chi}_{\mathrm{mm}}^{xx}&=2j\cdot\frac{t(0)-r(0)-1}{t(0)-r(0)-1}.
    \end{aligned}
\end{equation}

Our next goal is to characterize the normal susceptibilities, $\widetilde{\chi}_{\mathrm{mm}}^{zz}$ and $\widetilde{\chi}_{\mathrm{mm}}^{xz}=\widetilde{\chi}_{\mathrm{mm}}^{zx}$, given the already-extracted tangential susceptibilities $\widetilde{\chi}_{\mathrm{ee}}^{yy}$ and $\widetilde{\chi}_{\mathrm{mm}}^{xx}$. We therefore need to retrieve the full-wave reflection and transmission coefficients, $r(k_{x})$ and $t(k_{x})$, for at least one additional oblique angle of incidence $\theta_0$ (one nonzero tangential wavenumber $\widetilde{k}_{x}=\sin\theta_0$). Substituting these simulated values in Eq.\ (\ref{Eq:ScatCoeffs}) will again provide us with two equations relating between the two unknown normal susceptibilities. However, in this scenario, these equations depend quadratically on $\widetilde{\chi}_{\mathrm{mm}}^{xz}$ [see the expressions for $d_{1}$, $d_{3}$, and $t_{2}$ in Eqs.\ (\ref{Eq:d_coeffs}) and (\ref{Eq:t_coeffs})], which generally leads to two possible solutions, that is, ambiguity. This is not altogether surprising, because, as discussed in the main text, the presence of the $\widetilde{\chi}_{\mathrm{mm}}^{xz}$ component provides a distinct symmetry breaking, $t(-k_{x})\neq t(k_{x})$, the information of which is not sufficiently conveyed from the scattering at normal incidence and a single oblique angle alone. To exhaust all the underlying degrees of freedom in the system and obtain a unique set of susceptibilities, we will therefore need an additional simulated set of scattering coefficients at another oblique angle of incidence.

Before designating the second angle, let us derive more convenient relations for the oblique incidence scenario of the first angle $\theta_0$. Given the simulated values $r(k_{x})$ and $t(k_{x})$, the wavenumbers $\widetilde{k}_{x}=\sin\theta_0$ and $\widetilde{k}_{z}=\cos\theta_0$, and the known tangential susceptibilities [$\widetilde{\chi}_{\mathrm{ee}}^{yy}$ and $\widetilde{\chi}_{\mathrm{mm}}^{xx}$, Eq.\ (\ref{Eq:TangentialChar})], it is possible to combine both the equations in Eq.\ (\ref{Eq:ScatCoeffs}) and compose the expression
\begin{equation}
    \frac{t(k_{x})+1}{r(k_x)}=\frac{d_0-8j\widetilde{k}_{z}+jt_{1}\widetilde{k}_{z}\widetilde{k}_{x}+d_{2}\widetilde{k}_{z}^{2}}{r_0+r_{2}\widetilde{k}_{z}^{2}},
\end{equation}
from which we can extract a linear relation between $\widetilde{\chi}_{\mathrm{mm}}^{xz}$ and $\widetilde{\chi}_{\mathrm{mm}}^{zz}$,
\begin{equation}
\label{Eq:LinearRelationsPositive}
    \widetilde{\chi}_{\mathrm{mm}}^{xz}=-\frac{1}{4\widetilde{k}_{z}\widetilde{k}_{x}}\left[8j\widetilde{k}_{z}-d_0-d_{2}\widetilde{k}_{z}^{2}+\left(r_{0}+r_{2}\widetilde{k}_{z}^{2}\right)\frac{1+t(k_{x})}{r(k_{x})}\right]
\end{equation}
[the RHS depends linearly on $\widetilde{\chi}_{\mathrm{mm}}^{zz}$ via $r_{0}=-d_0=-2(\widetilde{\chi}_{\mathrm{ee}}^{yy}-\widetilde{\chi}_{\mathrm{mm}}^{zz})$ and $d_{2}=2(\widetilde{\chi}_{\mathrm{mm}}^{xx}-\widetilde{\chi}_{\mathrm{mm}}^{zz})$, Eqs.\ (\ref{Eq:d_coeffs}) and (\ref{Eq:r_coeffs})]. Next, let us also simulate for the scattering coefficients $t(-k_{x})$ and $r(-k_{x})$ at the opposite angle $-\theta_0$, for which the transverse wavenumber is $-\widetilde{k}_{x}=-\sin\theta_0$ and the normal wavenumber remains $\widetilde{k}_{z}=\cos\theta_0$. Furthermore, $r(-k_{x})=r(k_{x})$ from reciprocity (see Sec.\ \ref{Subsec:GeneralASymmetric}). We thus obtain another linear relation between $\widetilde{\chi}_{\mathrm{mm}}^{xz}$ and $\widetilde{\chi}_{\mathrm{mm}}^{zz}$,
\begin{equation}
\label{Eq:LinearRelationsNegative}
    \widetilde{\chi}_{\mathrm{mm}}^{xz}=\frac{1}{4\widetilde{k}_{z}\widetilde{k}_{x}}\left[8j\widetilde{k}_{z}-d_0-d_{2}\widetilde{k}_{z}^{2}+\left(r_{0}+r_{2}\widetilde{k}_{z}^{2}\right)\frac{1+t(-k_{x})}{r(k_{x})}\right].
\end{equation}
We finally equate Eqs.\ (\ref{Eq:LinearRelationsPositive}) and (\ref{Eq:LinearRelationsNegative}) and obtain a linear equation for $\widetilde{\chi}_{\mathrm{mm}}^{zz}$, whose solution is
\begin{equation}
\label{Eq:Char_mm_zz}
    \widetilde{\chi}_{\mathrm{mm}}^{zz}=-\frac{\widetilde{\chi}_{\mathrm{ee}}^{yy}}{\widetilde{k}_{x}^{2}}+\frac{8j\widetilde{k}_{z}r(k_{x})+\widetilde{\chi}_{\mathrm{mm}}^{xx}\widetilde{k}_{z}^{2}\left[2+t(k_{x})+t(-k_{x})-2r(k_{x})\right]}{\widetilde{k}_{x}^{2}\left[2+t(k_{x})+t(-k_{x})+2r(k_{x})\right]}.
\end{equation}

We can thus extract the associated $\widetilde{\chi}_{\mathrm{mm}}^{zz}$ value by substituting the already-extracted tangential susceptibilities [Eq.\ (\ref{Eq:TangentialChar})] and the simulated scattering coefficients at $\theta_0$ [$t(k_{x})$ and $r(k_{x})$] and $-\theta_0$ [$t(-k_{x})$ and $r(-k_{x})=r(k_{x})$] in Eq.\ (\ref{Eq:Char_mm_zz}). Then, this value can be used to evaluate $r_{0}=-d_0=-2(\widetilde{\chi}_{\mathrm{ee}}^{yy}-\widetilde{\chi}_{\mathrm{mm}}^{zz})$ and $d_{2}=2(\widetilde{\chi}_{\mathrm{mm}}^{xx}-\widetilde{\chi}_{\mathrm{mm}}^{zz})$ [Eqs.\ (\ref{Eq:d_coeffs}) and (\ref{Eq:r_coeffs})], which, in turn, can be substituted in Eq.\ (\ref{Eq:LinearRelationsPositive}) or in Eq.\ (\ref{Eq:LinearRelationsNegative}) to extract $\widetilde{\chi}_{\mathrm{mm}}^{xz}=\widetilde{\chi}_{\mathrm{mm}}^{zx}$. This concludes our extended characterization methodology. Note that in the particular scenario where the asymmetric component $\widetilde{\chi}_{\mathrm{mm}}^{xz}$ vanishes, then $t(-k_{x})=t(k_{x})$ (i.e., only one oblique incidence scenario at $\theta_0$ is required) and it can be shown that the expression of $\widetilde{\chi}_{\mathrm{mm}}^{zz}$ reduces to the one in \cite{Holloway2009,Zaluski2016,Shaham2021,Shaham2024AdvOptMat}.
%
%

\begin{backmatter}
\bmsection{Funding} {This research is supported by the Israel Science Foundation (Grant 3282/25).}

\bmsection{Acknowledgment}
{A.\ S.\ gratefully acknowledges generous support from the Andrew and Erna Finci Viterbi PhD Student Fellowship for the 2024-2025 academic year, and from the Andrew and Erna Finci Viterbi Postdoctoral Fellowship for Nurturing Future Faculty Members for the 2025-2026 academic year, received from the Andrew and Erna Viterbi Faculty of Electrical and Computer Engineering at the Technion---Israel Institute of Technology.}

\bmsection{Disclosures}
The authors declare no conflicts of interest.

\bmsection{Data availability} Data underlying the results presented in this paper may be obtained from the authors upon reasonable request.


\end{backmatter}


\end{document}